\documentclass[aps,twocolumn,amsmath,amssymb,superscriptaddress,prl,showpacs]{revtex4}
\usepackage[dvipdfmx]{graphicx}
\usepackage[dvipdfmx]{color}
\usepackage{amsmath}
\usepackage{amssymb}
\usepackage{dcolumn}
\usepackage{bm}
\usepackage{latexsym} 
\usepackage{setspace}

\begin{document}

\title{Quantitative Temperature Dependence of Longitudinal Spin Seebeck Effect\\ at High Temperatures}

\author{Ken-ichi Uchida}
\email{kuchida@imr.tohoku.ac.jp}
\affiliation{Institute for Materials Research, Tohoku University, Sendai 980-8577, Japan}
\affiliation{PRESTO, Japan Science and Technology Agency, Saitama 332-0012, Japan}

\author{Takashi Kikkawa}
\affiliation{Institute for Materials Research, Tohoku University, Sendai 980-8577, Japan}

\author{Asuka Miura}
\affiliation{Department of Mechanical Engineering, The University of Tokyo, Tokyo 113-8656, Japan}

\author{Junichiro Shiomi}
\affiliation{PRESTO, Japan Science and Technology Agency, Saitama 332-0012, Japan}
\affiliation{Department of Mechanical Engineering, The University of Tokyo, Tokyo 113-8656, Japan}

\author{Eiji Saitoh}
\affiliation{Institute for Materials Research, Tohoku University, Sendai 980-8577, Japan}
\affiliation{WPI Advanced Institute for Materials Research, Tohoku University, Sendai 980-8577, Japan}
\affiliation{CREST, Japan Science and Technology Agency, Tokyo 102-0076, Japan}
\affiliation{Advanced Science Research Center, Japan Atomic Energy Agency, Tokai 319-1195, Japan}
\date{\today}
\begin{abstract}
This article reports temperature-dependent measurements of longitudinal spin Seebeck effects (LSSEs) in Pt/Y$_3$Fe$_5$O$_{12}$ (YIG)/Pt systems in a high temperature range from room temperature to above the Curie temperature of YIG. The experimental results show that the magnitude of the LSSE voltage in the Pt/YIG/Pt systems rapidly decreases with increasing the temperature and disappears above the Curie temperature. The critical exponent of the LSSE voltage in the Pt/YIG/Pt systems at the Curie temperature was estimated to be 3, which is much greater than that for the magnetization curve of YIG. This difference highlights the fact that the mechanism of the LSSE cannot be explained in terms of simple static magnetic properties in YIG. \end{abstract}
\pacs{85.75.-d, 72.25.-b, 72.15.Jf, 73.50.Lw}
%
%
%
%
\maketitle
%
%
\section{I.~~~INTRODUCTION}
%
%
The Seebeck effect converts a temperature difference into electric voltage in conductors \cite{Seebeck1}. Since the discovery of the Seebeck effect nearly 200 years ago, it has been studied intensively to realize simple and environmentally-friendly energy-conversion technologies \cite{SeebeckApplication}. The Seebeck effect has been measured using various materials in a wide temperature range to investigate thermoelectric conversion performance and thermoelectric transport properties. Temperature-dependent measurements in a high temperature range are especially important in the investigation of the Seebeck effect, since thermoelectric devices are often used above room temperature \cite{SeebeckHighT1,SeebeckHighT2}. \par
In the field of spintronics \cite{spintronics1,spintronics2,spintronics3}, a spin counterpart of the Seebeck effect ---the spin Seebeck effect (SSE)--- was recently discovered. The SSE converts a temperature difference into spin voltage in ferromagnetic or ferrimagnetic materials. When a conductor is attached to a magnet under a temperature gradient, the thermally generated spin voltage in the magnet injects a spin current \cite{spincurrent1,spincurrent2,spincurrent3} into the conductor. Since the SSE occurs not only in metals \cite{SSE1,SSE10,SSE19,SSE27} and semiconductors \cite{SSE4,SSE9} but also in insulators \cite{SSE3,SSE5,SSE12,SSE13,SSE16,SSE17,SSE18,SSE19,SSE21,SSE24,SSE25,SSE26,SSE_Rezende,SSE_time_resolved1,SSE_time_resolved2,SSEJAPfull,SSE-JPCM}, it enables the construction of ``insulator-based thermoelectric generators'' \cite{SSE13} in combination with the inverse spin Hall effect (ISHE) \cite{ISHE1,ISHE2,ISHE3,ISHE4,ISHE5}, which was impossible if only the conventional Seebeck effect was used. \par
The observation of the SSE in insulators has been reported mainly in a longitudinal configuration \cite{SSE5,SSE12,SSEJAPfull,SSE13,SSE16,SSE17,SSE18,SSE19,SSE21,SSE24,SSE25,SSE26,SSE_Rezende,SSE_time_resolved1,SSE_time_resolved2,SSE-JPCM}. The sample system for measuring the longitudinal SSE (LSSE) is a simple paramagnetic metal (PM)/ferrimagnetic insulator (FI) junction system. In many cases, Pt and Y$_3$Fe$_5$O$_{12}$ (YIG) are used as PM and FI, respectively. When a temperature gradient $\nabla T$ is applied to the PM/FI system perpendicular to the interface, the spin voltage is thermally generated and injects a spin current into the PM along the $\nabla T$ direction owing to thermal spin-pumping mechanism \cite{SSE2,SSE6,SSE7,SSE8,SSE14,SSE15,SSE20,LSSE_Adachi,SSE23}. This thermally induced spin current is converted into an electric field ${\bf E}_{\rm ISHE}$ by the ISHE in the PM according to the relation
\begin{equation}\label{equ:SSE1}
{\bf E}_{\rm ISHE} \propto {\bf J}_{\rm s} \times {\bm \sigma}, 
\end{equation}
where ${\bf J}_{\rm s}$ is the spatial direction of the thermally induced spin current and ${\bm \sigma}$ is the spin-polarization vector of electrons in the PM, which is parallel to the magnetization ${\bf M}$ of FI [see Fig. \ref{fig:1}(a)]. By measuring ${\bf E}_{\rm ISHE}$ in the PM, one can detect the LSSE electrically. \par
In the experimental research on the SSE, temperature-dependent measurements also have been used for investigating various thermo-spin transport properties, such as phonon-mediated effects \cite{SSE4,SSE9,SSE6,SSE11}, correlation between the SSE and magnon excitation \cite{SSE_Rezende,Boona}, and effects of metal-insulator phase transition \cite{SSE19}. However, all the experiments on the SSE to date have been performed around and below room temperature. In this article, we report quantitative temperature-dependent measurements of the LSSE in Pt/YIG systems in the high temperature range from room temperature to above the Curie temperature of YIG. \par
%
%
%
\section{II.~~~EXPERIMENTAL PROCEDURE} \label{sec:procedure}
%
%
%
%
\begin{figure*}[tb]
\begin{center}
\includegraphics{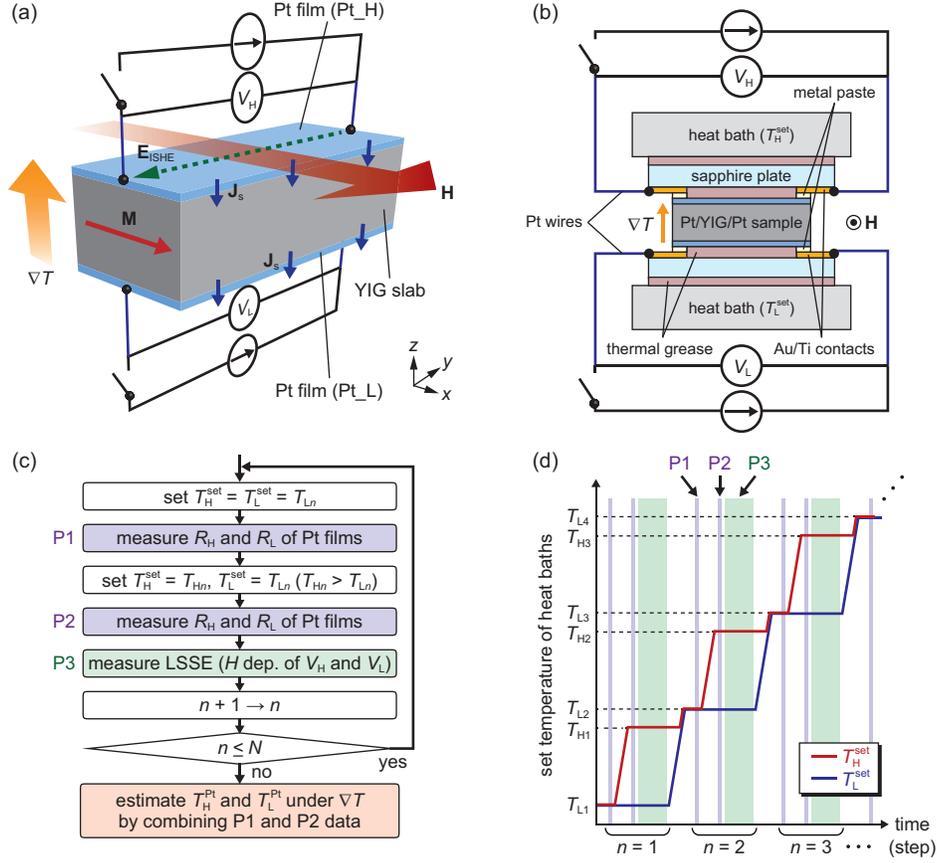}
\caption{(a) A schematic illustration of the Pt/YIG/Pt sample. $\nabla T$, ${\bf H}$, ${\bf M}$, ${\bf E}_{\rm ISHE}$, and ${\bf J}_{\rm s}$ denote the temperature gradient, magnetic field (with the magnitude $H$), magnetization vector, electric field induced by the ISHE, and spatial direction of the thermally generated spin current, respectively. The electric voltage $V_{\rm H}$ ($V_{\rm L}$) and resistance $R_{\rm H}$ ($R_{\rm L}$) between the ends of Pt$\_$H (Pt$\_$L) were measured using a nanovolt/micro-ohm meter (Agilent 34420A) in the `DC-voltage' and `2-wire-resistance' modes, respectively. The DC-voltage (2-wire-resistance) mode corresponds to the switch-off (switch-on) state in this schematic illustration. (b) Experimental configuration for measuring the LSSE used in the present study. (c) A flow chart of the measurement processes. (d) A schematic graph of the set temperatures, $T_{{\rm H}n}$ and $T_{{\rm L}n}$, of the heat baths as a function of time or the measurement step number $n$. The process P1 was performed under the isothermal condition, while the processes P2 and P3 were under the temperature gradient. Before starting the processes P1 and P2, we waited for $\sim 30~\textrm{minutes}$ at each $n$ to stabilize the $T_{\rm H}^{\rm set}$ and $T_{\rm L}^{\rm set}$ values. }\label{fig:1}
\end{center}
\end{figure*}
The sample system used in this study consists of a single-crystalline YIG slab covered with Pt films. One difference from conventional samples is that the Pt films are put on both the top and bottom surfaces of the YIG slab [Fig. \ref{fig:1}(a)], while only the top surface of YIG is covered with a Pt film in conventional samples \cite{SSE5,SSE17,SSE25,SSE-JPCM}. The lengths of the YIG slab along the $x$, $y$, and $z$ directions are 3 mm, 7 mm, and 1 mm, respectively. 10-nm-thick Pt films were sputtered on the whole of the $3 \times 7$-mm$^2$ (111) surfaces of the YIG. The top and bottom Pt films are electrically insulated from each other because YIG is a very good insulator. Since YIG has the large charge gap of 2.7 eV \cite{Kajiwara_nature}, thermal excitation of charge carriers in YIG is vanishingly small even at the high temperatures. \par
To attach electrodes to both the Pt films symmetrically and to generate a uniform temperature gradient, we made the configuration shown in Fig. \ref{fig:1}(b). Here, the Pt/YIG/Pt sample was sandwiched between two 0.5-mm-thick sapphire plates of which the surface is covered with two separated Au/Ti contacts. The distance between the two Au/Ti contacts is $\sim 6~\textrm{mm}$. To extract voltage signals in the Pt films, both the ends of the Pt films are connected to the Au/Ti contacts via sintering metal paste, which can be used up to 900 $^\circ$C, and thin Pt wires with the diameter of 0.1 mm were attached to the end of the contacts [see Fig. \ref{fig:1}(b)]. The sapphire plates are thermally connected to heat baths of which the temperatures are controlled with the accuracy of $<0.6~\textrm{K}$ by using PID (proportional-integral-derivative) temperature controllers. We attached thermal paste to both the surfaces of the sapphire plates, except for the regions of the Au/Ti contacts, to improve the thermal contact [see Fig. \ref{fig:1}(b)]. During the measurements of the LSSE, the temperature of the upper heat bath, $T_{\rm H}^{\rm set}$, is set to be higher than that of the lower one, $T_{\rm L}^{\rm set}$. According to the direction of the temperature gradient, hereafter, the top and bottom Pt films of the Pt/YIG/Pt sample are referred to as `Pt$\_$H' and `Pt$\_$L', respectively. In this setup, we can measure the electric voltage $V_{\rm H}$ ($V_{\rm L}$) and resistance $R_{\rm H}$ ($R_{\rm L}$) between the ends of Pt$\_$H (Pt$\_$L) without changing electrodes, wiring, and measurement equipment [see also the caption of Fig. \ref{fig:1}]. An external magnetic field ${\bf H}$ (with the magnitude $H$) was applied along the $x$ direction. \par
To investigate the temperature dependence of the LSSE quantitatively, it is important to estimate the temperature difference between the top and bottom of the YIG sample. However, in the conventional experiments, the temperatures of the heat baths, not the sample itself, were usually monitored \cite{SSEJAPfull}. Therefore, the measured temperature difference includes the contributions from the interfacial thermal resistance between the sample and heat baths and from small temperature gradients in the sample holders. To avoid this problem, in this study, we used the Pt films not only as spin-current detectors but also as temperature sensors \cite{SSE26,SSE_time_resolved2}; we can know the temperatures of the Pt films from the temperature dependence of the resistance of the films, enabling the estimation of the temperature difference between the top and bottom of the sample during the LSSE measurements. \par
\begin{table}[b]
\setlength{\belowcaptionskip}{0mm}
\caption{Set temperatures of the heat baths for each measurement step number $n~(\leq N)$. Here, the $T_{{\rm H}n}$ and $T_{{\rm L}n}$ values are increased by 10 K with every $n$ increase, while $T_{{\rm H}n}-T_{{\rm L}n}$ is fixed at 8 K. }
\label{tab:1}
\begin{center}
\begin{tabular}{ccc} \hline \hline
~~Step number $n$~~ & ~~$T_{{\rm L}n}$ (K)~~ & ~~$T_{{\rm H}n}$ (K)~~ \\\hline 
1 & 290 & 298 \\
2 & 300 & 308 \\
3 & 310 & 318 \\
4 & 320 & 328 \\
5 & 330 & 338 \\
$\cdot$ & $\cdot$ & $\cdot$ \\
$\cdot$ & $\cdot$ & $\cdot$ \\
$\cdot$ & $\cdot$ & $\cdot$ \\
31 & 590 & 598 \\
32 & 600 & 608 \\
33 & 610 & 618 \\
34 ($= N$) & 620 & 628 \\ \hline \hline
\end{tabular}
\end{center}
\end{table}
Figure \ref{fig:1}(c) shows the flow chart of the measurement processes. The measurement comprises the following three processes P1-P3, and the processes are repeated $N$ times. Hereafter, $T_{{\rm H}n}$ and $T_{{\rm L}n}$ denote the set temperatures for each measurement step number $n~(=1,~2,...,~N)$ (see Table \ref{tab:1}, where the values of $T_{{\rm H}n}$ and $T_{{\rm L}n}$ for each $n$ are shown). The process P1 is the measurement of the resistance of Pt$\_$H and Pt$\_$L, $R_{\rm H}$ and $R_{\rm L}$, in the isothermal condition, where the temperatures of the heat baths are set to the same temperature: $T_{\rm H}^{\rm set}=T_{\rm L}^{\rm set} = T_{{\rm L}n}$. In this isothermal condition, the temperature of the Pt/YIG/Pt sample is uniform and very close to the set temperature of the heat baths irrespective of the presence of the interfacial thermal resistance. Next, we apply a temperature gradient to the Pt/YIG/Pt sample by increasing the temperature of the upper heat bath, where $T_{\rm H}^{\rm set}=T_{{\rm H}n}$ and $T_{\rm L}^{\rm set}=T_{{\rm L}n}$ with $T_{{\rm H}n} > T_{{\rm L}n}$ (see Table \ref{tab:1}). After waiting until the temperatures are stabilized, we measure the resistance of the Pt films under the temperature gradient: this is the process P2. The processes P1 and P2 are performed without applying ${\bf H}$. Immediately after the process P2, we proceed to the process P3; the LSSE, i.e. the $H$ dependence of $V_{\rm H}$ and $V_{\rm L}$, is measured with keeping the magnitude of $T_{\rm H}^{\rm set} - T_{\rm L}^{\rm set}$ constant. After finishing the LSSE measurements, we go on to the next step and increase $n$ by 1, where the values of $T_{{\rm H}n}$ and $T_{{\rm L}n}$ are increased as shown in Table \ref{tab:1}. These measurement processes are summarized in Fig. \ref{fig:1}(d). \par
The calibration method of the sample temperatures is as follows. From the process P1, we obtain the temperature ($T_{{\rm L}n}$) dependence of $R_{\rm H}$ and $R_{\rm L}$ under the isothermal condition. By comparing the resistance under the temperature gradient, obtained from the process P2, with the isothermal $R_{\rm H,L}$-$T_{{\rm L}n}$ curves, we can calibrate the temperature of the Pt films, $T_{\rm H}^{\rm Pt}$ and $T_{\rm L}^{\rm Pt}$, under the temperature gradient, allowing us to estimate the average temperature $T_{\rm av}$ ($=(T_{\rm H}^{\rm Pt}+T_{\rm L}^{\rm Pt})/2$) and the temperature difference $\Delta T$ ($=T_{\rm H}^{\rm Pt}-T_{\rm L}^{\rm Pt}$) in the YIG slab during the LSSE measurements. The  $T_{\rm av}$ and $\Delta T$ values are free from the contributions from the interfacial thermal resistance between the sample and heat baths and from the temperature gradients in the sapphire plates, enabling the quantitative evaluation of the LSSE at various temperatures. Although the experimental protocol proposed here cannot estimate the contributions of the interfacial thermal resistance at the Pt/YIG interfaces and temperature gradients in the Pt films, they are negligible compared with the temperature difference applied to the YIG slab \cite{SSE24,SSE-JPCM}. \par
%
%
\section{III.~~~RESULTS AND DISCUSSION}
%
%
\begin{figure*}[tb]
\begin{center}
\includegraphics{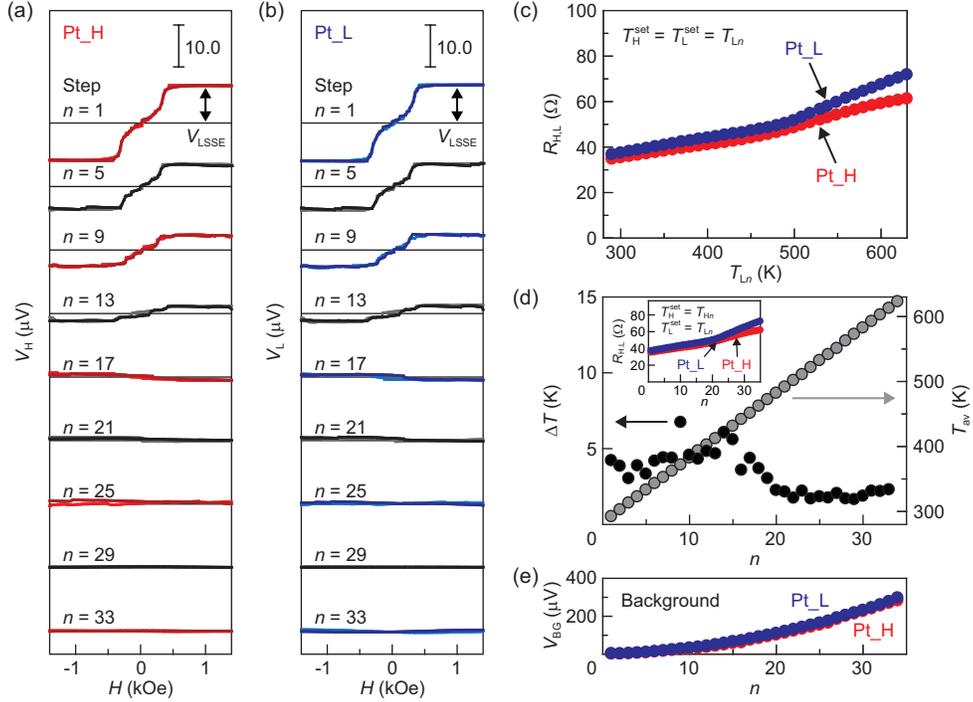}
\caption{(a) $H$ dependence of $V_{\rm H}$ for Pt$\_$H in the Pt/YIG/Pt sample for various values of $n$. (b) $H$ dependence of $V_{\rm L}$ for Pt$\_$L. The LSSE voltage $V_{\rm LSSE}$ for Pt$\_$H (Pt$\_$L) is defined as $V_{\rm H}$ ($V_{\rm L}$) at $H=1~\textrm{kOe}$. (c) $T_{{\rm L}n}$ dependence of $R_{\rm H}$ ($R_{\rm L}$) for Pt$\_$H (Pt$\_$L) in the isothermal condition: $T_{\rm H}^{\rm set}=T_{\rm L}^{\rm set} = T_{{\rm L}n}$. (d) The average temperature $T_{\rm av}$ and temperature difference $\Delta T$ of the Pt/YIG/Pt sample during the LSSE measurements as a function of $n$. The inset to (d) shows $R_{\rm H}$ ($R_{\rm L}$) for Pt$\_$H (Pt$\_$L) as a function of $n$ under the temperature gradient: $T_{\rm H}^{\rm set}=T_{{\rm H}n}$ and $T_{\rm L}^{\rm set}=T_{{\rm L}n}$ with $T_{{\rm H}n} > T_{{\rm L}n}$. (e) The background voltage $V_{\rm BG}$ for Pt$\_$H and Pt$\_$L as a function of $n$ under the temperature gradient. }\label{fig:2}
\end{center}
\end{figure*}
Figures \ref{fig:2}(a) and \ref{fig:2}(b) respectively show the $H$ dependence of $V_{\rm H}$ and $V_{\rm L}$ in the Pt/YIG/Pt sample for each step number $n$, measured when $T_{{\rm H}n}-T_{{\rm L}n} = 8~\textrm{K}$. Around room temperature, we observed clear voltage signals in both the Pt films; the signs of $V_{\rm H}$ and $V_{\rm L}$ are reversed in response to the reversal of the magnetization direction of the YIG slab. Since the contribution of anomalous Nernst effects induced by proximity ferromagnetism in Pt \cite{Proximity} is negligibly small in Pt/YIG systems, the voltage signals observed here are due purely to the LSSE \cite{SSE17,SSE25,SSE-JPCM}. The sign of the LSSE voltage in Pt$\_$H was observed to be the same as that in Pt$\_$L, a situation consistent with the scenario of the SSE \cite{SSE2,SSE7,SSE20,comment_sign}. Here we note that, although the large background voltage $V_{\rm BG}$ appears due to unavoidable thermopower differences in the wires with increasing $n$ [Fig. \ref{fig:2}(e)], the noise level in the $V_{\rm H}$ and $V_{\rm L}$ signals does not increase and the drift of $V_{\rm BG}$ is small [Figs. \ref{fig:2}(a) and \ref{fig:2}(b)]. Therefore, we can extract the LSSE voltage simply by measuring the $H$ dependence of $V_{\rm H}$ and $V_{\rm L}$. We also found that the magnitude of the LSSE voltage in the Pt/YIG/Pt sample monotonically decreases with increasing the temperature. \par
To quantitatively discuss the temperature dependence of the LSSE voltage in the Pt/YIG/Pt sample, we estimate $T_{\rm av}$ and $\Delta T$ at each step number $n$ by using the method explained in Sec. II. Figure \ref{fig:2}(c) shows the $T_{{\rm L}n}$ dependence of $R_{\rm H}$ and $R_{\rm L}$ for the Pt films measured under the isothermal condition. By combining the isothermal $R_{\rm H,L}$-$T_{{\rm L}n}$ curves with the $R_{\rm H}$ and $R_{\rm L}$ data under the temperature gradient [the inset to Fig. \ref{fig:2}(d)], we obtain the $T_{\rm av}$ and $\Delta T$ values at each $n$ [Fig. \ref{fig:2}(d)]. Importantly, the calibrated values of $\Delta T$ are dependent on $n$ and always smaller than the temperature difference applied to the heat baths due to the interfacial thermal resistance and temperature gradients in the sapphire plates (note that $T_{{\rm H}n}-T_{{\rm L}n} = 8~\textrm{K}$ for all the measurements as shown in Table \ref{tab:1}). \par
\begin{figure*}[tb]
\begin{center}
\includegraphics{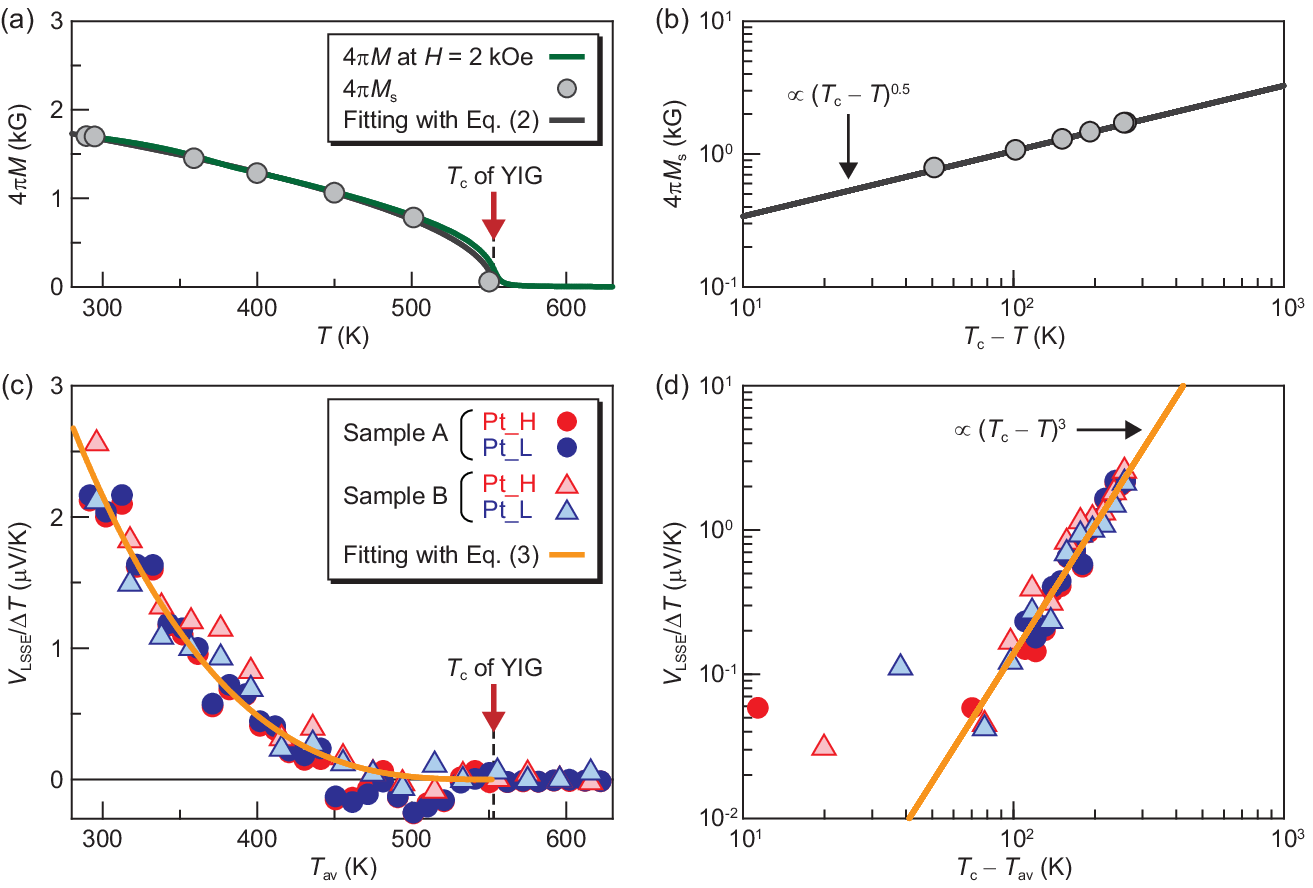}
\caption{(a) $T$ dependence of the bulk magnetization $4 \pi M$ of the YIG slab. The green curve shows the $4 \pi M$ data at $H=2~\textrm{kOe}$, measured with a vibrating sample magnetometer (VSM). The gray circles show the saturation magnetization $4 \pi M_{\rm s}$ of the YIG slab. The gray curve was obtained by fitting the $4 \pi M_{\rm s}$ data with Eq. (\ref{equ:critical1}). The values of $4 \pi M_{\rm s}$ and Curie temperature $T_{\rm c}~(=553~\textrm{K})$ of the YIG slab were estimated by using the Arrott plot method (see Appendix A). (b) Double logarithmic plot of the $T_{\rm c}-T$ dependence of $4 \pi M_{\rm s}$ of the YIG slab. (c) $T_{\rm av}$ dependence of $V_{\rm LSSE}/\Delta T$ for two different Pt/YIG/Pt samples A (circles) and B (triangles). The experimental results shown in Fig. \ref{fig:2} were measured using the Pt/YIG/Pt sample A. The orange curve was obtained by fitting the $V_{\rm LSSE}/\Delta T$ data with Eq. (\ref{equ:critical2}). (d) Double logarithmic plot of the $T_{\rm c}-T_{\rm av}$ dependence of $V_{\rm LSSE}/\Delta T$ for the Pt/YIG/Pt samples A and B. }\label{fig:3}
\end{center}
\end{figure*}
Figure \ref{fig:3}(c) shows the LSSE voltage normalized by the calibrated temperature difference applied to the Pt/YIG/Pt sample, $V_{\rm LSSE}/\Delta T$, as a function of $T_{\rm av}$. Here, $V_{\rm LSSE}$ denotes $V_{\rm H}$ ($V_{\rm L}$) for Pt$\_$H (Pt$\_$L) at $H=1~\textrm{kOe}$. We confirm again that the magnitude of $V_{\rm LSSE}/\Delta T$ monotonically decreases with increasing the temperature and disappears above the Curie temperature $T_{\rm c}$ of YIG, where $T_{\rm c}$ of our YIG slab was experimentally estimated to be 553 K (see Appendix A). This behavior was observed not only in one sample but also in our different samples as exemplified in Fig. \ref{fig:3}(c). Interestingly, the temperature dependence of $V_{\rm LSSE}/\Delta T$ is significantly different from the magnetization ($4\pi M$) curve of the YIG slab [compare Figs. \ref{fig:3}(a) and \ref{fig:3}(c)]; the magnitude of $V_{\rm LSSE}/\Delta T$ rapidly decreases with a concave-up shape, while the magnetization curve of YIG exhibits a standard concave-down shape. We also checked that the strong temperature dependence of the LSSE voltage in the Pt/YIG/Pt sample cannot be explained by the weak temperature dependence of the thermal conductivity of YIG (see Appendix B). Similar difference in the temperature-dependent data between the ISHE voltage and magnetization was observed also in Pt/GaMnAs systems in the measurement of the transverse SSEs \cite{SSE4,SSE9}. \par
The behavior of physical quantities near continuous phase transitions can be described by critical exponents in general. Here, we compare the critical exponents for the observed temperature dependences of the LSSE voltage in the Pt/YIG/Pt sample and the magnetization of YIG. First, we checked that the magnetization curve of YIG is well reproduced by a standard mean-field model \cite{Chikazumi}: 
\begin{equation}\label{equ:critical1}
4 \pi M_{\rm s} = A (T_{\rm c} - T)^{0.5}, 
\end{equation}
where the critical exponent is fixed at 0.5 and $A$ is an adjustable parameter [see Figs. \ref{fig:3}(a) and \ref{fig:3}(b)]. The critical exponent $\gamma$ for the LSSE was estimated by fitting the experimental data in Fig. \ref{fig:3}(c) with the following equation: 
\begin{equation}\label{equ:critical2}
\frac{{V_{\rm ISHE}}}{{\Delta T}} = S (T_{\rm c} - T)^\gamma, 
\end{equation}
where both $S$ and $\gamma$ are adjustable parameters and $T_{\rm av}$ is regarded as $T$ for the LSSE data. We found that the observed temperature dependence of $V_{\rm LSSE}/\Delta T$ for the Pt/YIG/Pt sample is well fitted by Eq. (\ref{equ:critical2}) with $\gamma = 3$, which is much greater than the critical exponent for the magnetization curve [see also the double logarithmic plot in Fig. \ref{fig:3}(d)]. This big difference in the critical exponents between the LSSE and magnetization emphasizes the fact that the LSSE is not attributed solely to static magnetic properties in YIG. \par
Here, we qualitatively discuss the origin of the temperature dependence of the LSSE voltage in the Pt/YIG/Pt sample. According to the thermal spin-pumping mechanism \cite{SSE2,SSE20} and phenomenological calculation of the ISHE combined with short-circuit effects \cite{NakayamaPRB}, the magnitude of the LSSE voltage is determined mainly by the following factors: the spin-mixing conductance \cite{Tserkovnyak05,Jia11a,Weiler13,Qiu13} at the Pt/YIG interfaces, spin-diffusion length and spin-Hall angle of Pt, and difference between an effective magnon temperature in YIG and an effective electron temperature in Pt. Since the LSSE voltage is proportional to the spin-mixing conductance \cite{SSE2}, it can contribute directly to the observed temperature dependence of the LSSE voltage. Recently, Ohnuma {\it et al.} formulated the relation between the spin-mixing conductance and interface $s$-$d$ interaction at paramagnet/ferromagnet interfaces, and predicted that the spin-mixing conductance is proportional to $(4 \pi M_{\rm s})^2$ of the ferromagnet \cite{Ohnuma2014}. By combining this prediction with Eq. (\ref{equ:critical1}), the spin-mixing conductance is proportional to $T_{\rm c} - T$, of which the critical exponent ($=1$) is greater than that for the magnetization curve. Although the temperature dependence of the spin-mixing conductance can explain the facts that the LSSE voltage monotonically decreases with increasing the temperature and disappears at $T_{\rm c}$, it is still much weaker than the observed $(T_{\rm c} - T)^{3}$ dependence of the LSSE voltage. Furthermore, if the spin-diffusion length of Pt decreases with increasing the temperature \cite{Marmion2014PRB}, it can also contribute to reducing the LSSE voltage at high temperatures, while the spin-Hall angle of Pt was shown to exhibit weak temperature dependence \cite{Vila2007} (note that the magnitude of the ISHE voltage monotonically decreases with decreasing the spin-diffusion length when the spin-Hall angle is constant \cite{NakayamaPRB}). The effective magnon-electron temperature difference could also be an important factor, but there is no clear framework to determine its temperature dependence at the present stage. Therefore, more elaborate investigations are necessary for the complete understanding of the temperature dependence of the LSSE voltage. \par
%
%
\section{IV.~~~CONCLUSION}
%
%
In this study, we reported the longitudinal spin Seebeck effects (LSSEs) in Y$_3$Fe$_5$O$_{12}$ (YIG) slabs sandwiched by two Pt films in the high temperature range from room temperature to above the Curie temperature $T_{\rm c}$ of YIG. To investigate the temperature dependence of the LSSE quantitatively, we used the Pt films not only as spin-current detectors but also as temperature sensors. The measurement processes used here enabled accurate estimation of the average temperature and temperature difference of the sample, being free from thermal artifacts. We found that the magnitude of the LSSE in the Pt/YIG/Pt sample rapidly decreases with increasing the temperature and disappears above $T_{\rm c}$ of YIG; the observed LSSE voltage exhibits the $(T_{\rm c} - T)^{3}$ dependence of which the critical exponent ($=3$) is much greater than that of the magnetization of YIG ($=0.5$). Although more detailed experimental and theoretical investigations are required to clarify the microscopic origin of this discrepancy, we anticipate that the quantitative temperature-dependent LSSE data at high temperatures will be helpful for obtaining full understanding of the mechanism of the LSSE. \par
%
%
\section*{ACKNOWLEDGMENTS}
%
The authors thank S. Maekawa, H. Adachi, Y. Ohnuma, N. Yokoi, K. Sato, and J. Ohe for valuable discussions and Y. Zhang for his assistance in magnetometry measurements. This work was supported by PRESTO-JST ``Phase Interfaces for Highly Efficient Energy Utilization'', CREST-JST ``Creation of Nanosystems with Novel Functions through Process Integration'', Grant-in-Aid for Young Scientists (A) (25707029), Grant-in-Aid for Challenging Exploratory Research (26600067), Grant-in-Aid for Scientific Research (A) (24244051) from MEXT, Japan, LC-IMR of Tohoku University, the Sumitomo Foundation, the Tanikawa Fund Promotion of Thermal Technology, the Casio Science Promotion Foundation, and the Iwatani Naoji Foundation. \par
%
%
\section*{APPENDIX A: ESTIMATION OF CURIE TEMPERATURE OF YIG}
\begin{figure}[tb]
\begin{center}
\includegraphics{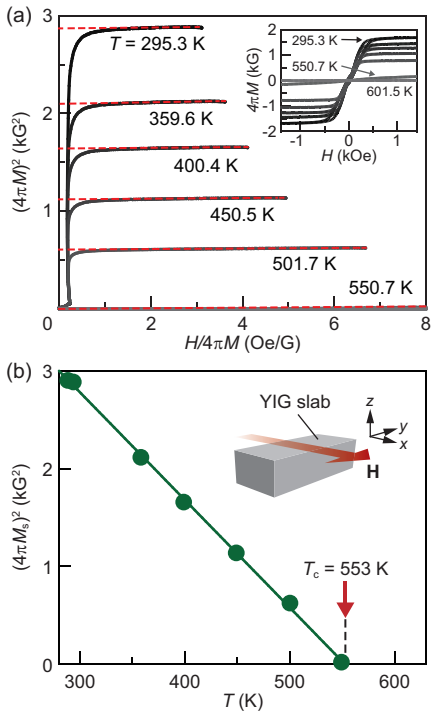}
\caption{(a) Arrott plots of the YIG slab for various values of $T$. The inset to (a) shows the $H$ dependence of $4 \pi M$ of the YIG slab for various values of $T$, measured with VSM. The lengths along the $x$, $y$, and $z$ directions of the YIG slab used for the magnetometry measurements are 3 mm, 7 mm, and 1 mm, respectively. ${\bf H}$ was applied along the $x$ direction. (b) $T$ dependence of $(4 \pi M_{\rm s})^2$ of the YIG slab. }\label{fig:4}
\end{center}
\end{figure}

%
%
The Curie temperature $T_{\rm c}$ of the YIG slab used in the present study was estimated from vibrating sample magnetometry and Arrott-plot analysis \cite{Chikazumi,Arrott}. The inset to Fig. \ref{fig:4}(a) shows the $H$ dependence of the magnetization $4 \pi M$ of the YIG slab for various values of $T$. From this result, we obtained the Arrott plots, i.e. $H/4 \pi M$ dependence of $(4 \pi M)^2$, of the YIG slab [see Fig. \ref{fig:4}(a)]. The saturation magnetization $4 \pi M_{\rm s}$ of the YIG at each temperature was extracted by extrapolating the $(4 \pi M)^2$ data in the high-magnetic-field range to zero field [see red dotted lines in Fig. \ref{fig:4}(a)]. As shown in Fig. \ref{fig:4}(b), the $T$ dependence of $(4 \pi M_{\rm s})^2$ of the YIG slab is well fitted by a linear function; the horizontal intercept of the linear fit line corresponds to $T_{\rm c}$. The fitting result shows that the Curie temperature of our YIG slab is $T_{\rm c}=553~\textrm{K}$, which is consistent with literature values \cite{YIG1,YIG2}. \par
%
%
%
\section*{APPENDIX B: TEMPERATURE DEPENDENCE OF THERMAL CONDUCTIVITY OF YIG}
\begin{figure}[tb]
\begin{center}
\includegraphics{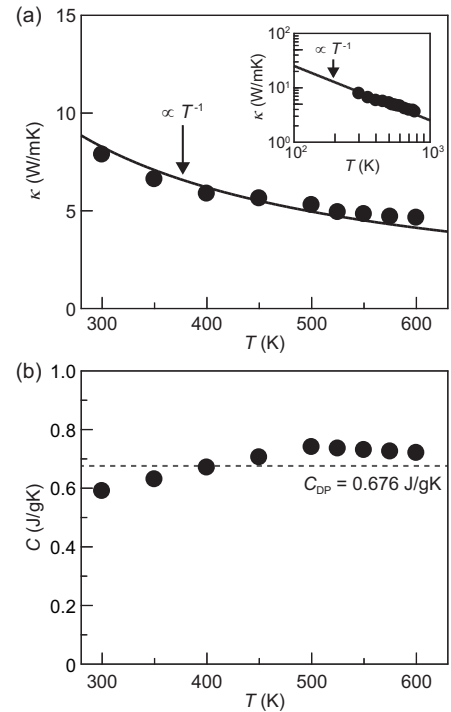}
\caption{(a) $T$ dependence of the thermal conductivity $\kappa$ of the YIG slab. The black circles show the measured thermal conductivity and the black curve shows the fitting of the experimental data with $\kappa = KT^{-1}$, where $K$ is an adjustable parameter. The inset shows the double logarithmic plot of the $T$ dependence of $\kappa$. (b) $T$ dependence of the specific heat $C$ of the YIG slab. The gray dotted line shows the DP specific heat of YIG. }\label{fig:5}
\end{center}
\end{figure}

%
%
Figure \ref{fig:5}(a) shows the thermal conductivity $\kappa$ of the YIG slab used in the present study as a function of $T$. The $\kappa$ values were obtained by the combination of thermal diffusivity measured by a laser-flash method and specific heat $C$ measured by a differential scanning calorimetry. Here, we measured the thermal diffusivity along the [111] direction of the single-crystalline YIG slab, which is parallel to the $\nabla T$ direction in the LSSE setup. As shown in Fig. \ref{fig:5}(b), the measured $C$ values are consistent with the Dulong-Petit (DP) law \cite{Seebeck1}; the difference of the $C$ values from the DP specific heat of YIG, $C_{\rm DP} = 0.676~\textrm{J/gK}$, is less than 10 \% of $C_{\rm DP}$ for $T > 350~\textrm{K}$. The observed $T$ dependence of $\kappa$ is well fitted by $\kappa \propto T^{-1}$, indicating that the thermal conductivity of the YIG is dominated by phonons in this temperature range [see also the inset to Fig. \ref{fig:5}(a)]. The $\kappa$ value at 300 K is consistent with literature values \cite{YIG_kappa1,YIG_kappa2,YIG_kappa3}. We also confirmed that the $T$ dependence of $\kappa$ is much weaker than that of the LSSE voltage in the Pt/YIG/Pt sample [compare Figs. \ref{fig:3}(c) and \ref{fig:5}(a)] and shows no anomaly around $T_{\rm c}$. 
\end{document}